\documentstyle[sprocl,epsfig]{article}

\def\Journal#1#2#3#4{{#1} {\bf #2}, #3 (#4)}

\def\CPC{\em Comput. Phys. Commun.}

\def\PLB{{\em Phys. Lett.}  B}
\def\PRL{\em Phys. Rev. Lett.}

\def\ZPC{{\em Z. Phys.} C}
\def\JPG{{\em J. Phys.} G}
\def\EPJ{{\em Eur. Phys. J.} C}

\begin{document}

\title{SOFT REMNANT INTERACTIONS AND RAPIDITY GAPS\footnote[1]{Presented by N.\ Timneanu at the DIS2000 Workshop, Liverpool, England, April 2000}}

\author{NICUSOR TIMNEANU and GUNNAR INGELMAN}

\address{High Energy Physics, Uppsala University, Box 535, S-75121, Uppsala, Sweden\\E-mail: nicusor@tsl.uu.se} 

\maketitle\abstracts{
Soft colour exchange models give a unified description of both diffractive and non-diffractive events, such that $ep$ and $p\bar{p}$ collider data with 
and without rapidity gaps are well reproduced. We show that these models
also describe the new Tevatron data on diffraction based on observed 
leading antiprotons, which provide new information on how to treat the 
beam particle remnant in the Monte Carlo model.
}

\vspace*{-75mm}
\noindent
TSL/ISV-2000-0230
\\June 2000    
\vspace*{64mm}

The hard scale in diffractive hard scattering~\cite{ingelman}
provides a parton level basis for investigating the non-perturbative 
effects on a long space-time scale that are important in diffraction. 
Traditionally, this has been explained in the Regge framework by the 
exchange of a pomeron with a parton structure. By fitting parton density functions in the pomeron, this gives a good description of diffractive 
deep inelastic scattering at HERA. However, applying exactly the same model 
for $p\bar{p}$ gives a too large cross section for diffractive dijet and $W$ production compared to Tevatron data~\cite{cdf-old}. 
This indicates a non-universality problem of the pomeron model, e.g.\ related to the pomeron flux. Since this flux specifies the leading particle spectrum, it is interesting to note that the new Tevatron data~\cite{cdf-ap} with a leading antiproton show a similar problem of the pomeron model.

A different approach is represented by the Soft Colour Interaction~\cite{SCI} (SCI) model and the Generalized Area Law~\cite{GAL} (GAL) model which consider 
non-perturbative interactions and their resulting variations in the topology of the confining colour force fields. The SCI model is based on the assumption that partons emerging from the hard interaction can exchange colour with the proton colour field in which they propagate. In GAL, the interaction between overlapping strings is the basis for colour exchange. Both models lead to rearrangement of the string topology, in the former related to a fixed probability to exchange a soft gluon between pairs of partons, in the latter with a probability depending on string dynamics. 

Both models are implemented in the Monte Carlo generators {\sc Lepto}~\cite{LEPTO} for $ep$ and {\sc Pythia}~\cite{PYTHIA} for $p\bar{p}$. Using standard Lund model hadronisation, different string topologies lead to different hadronic final states; with or without rapidity gaps, leading particles etc. This unified description of all final states gives a good description of HERA data on rapidity gap events~\cite{SCI} and leading neutrons~\cite{hera-ln}. 
The latter is not well described by single pion exchange in the Regge approach. 

\begin{figure}[th]
\vspace*{-5mm}
\center{\epsfig{width=0.95 \columnwidth,file=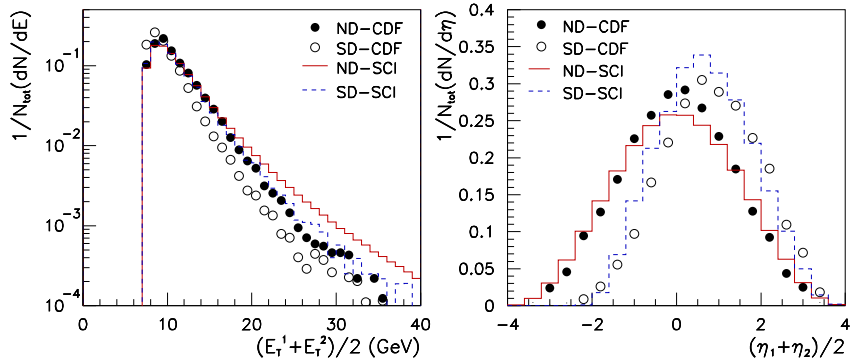,
bbllx=5, bblly=10, bburx=560, bbury=250}}
\vspace*{-3mm}
\caption{Distributions of mean transverse energy and pseudorapidity of the dijets in non-diffractive (ND) and single diffractive (SD) $p\bar{p}$ events. Points are Tevatron (CDF) data \protect\cite{cdf-ap} and histograms are standard {\sc Pythia} with added soft colour interactions (SCI). 
\label{fig:1}}
\vspace*{-3mm}
\end{figure}

The same SCI and GAL models also describe well~\cite{durham} the production of W and dijets with rapidity gaps observed~\cite{cdf-old} in $p\bar{p}$ at the Tevatron. Here we report the comparison of our models with the new Tevatron data~\cite{cdf-ap} on dijet events with an observed leading antiproton which signals diffraction. For this event sample, the observed $E_T$ and $\eta$ distribution of the jets are reasonably well described by the models (Fig.~1). 
In both data and models the jet-$E_T$ distribution of the single diffractive events is somewhat steeper than in the inclusive non-diffractive sample. The somewhat steeper overall slope in data compared to the model may be due to a mismatch in the jet reconstruction or the lack of NLO corrections in {\sc Pythia}'s hard scattering matrix elements.

The ratio of diffractive to non-diffractive dijet events as a function of momentum fraction of the struck parton in the antiproton is shown in Fig.~2a.
The pomeron model (tuned to HERA diffractive data) is far above the 
CDF data, while default {\sc Pythia} is far below. The soft colour exchange models give a quite reasonable description of the data. However, 
the agreement is not perfect since the slope of the CDF data remains constant 
when the data is split in bins of different $\xi$ ($\xi=1-x_F$), whereas the SCI model then gives a somewhat varying slope. This relates to the details of how
the beam particle remnant is treated in the model, which is not known from first principles. In order to set up the string connections the hadron remnant usually needs to be split in two parts taking energy fractions $\chi$ and $1-\chi$ given by a parametrisation ${\cal P}(\chi)$. These can be taken from simple counting rule arguments ($\sim (1-\chi)$) or as a parton distribution ($\sim \chi^{-1} (1-\chi)^3$). This gives a substantial model uncertainty in the Feynman-$x$ distribution of the leading $\bar{p}$ as shown in Fig.~2b.

\begin{figure}[th]
\vspace*{-5mm}
\center{\epsfig{width=0.95 \columnwidth,file=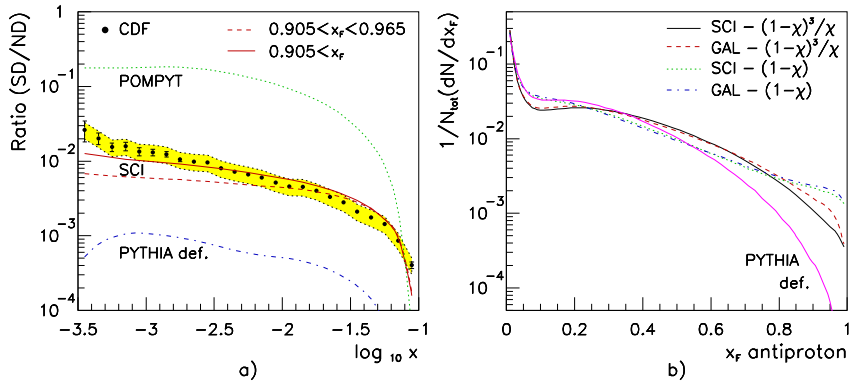,
bbllx=10, bblly=0, bburx=560, bbury=250}}
\vspace*{-3mm}
\caption{{\bf(a)} Ratio of diffractive to non-diffractive dijet events versus  momentum fraction $x$ of the interacting parton in $\bar{p}$.  Points are CDF data, with a systematic normalization uncertainty of $\pm 25 \%$ and curves 
are from the {\sc Pompyt} pomeron model, default {\sc Pythia} and {\sc Pythia} with soft colour interactions (SCI). {\bf(b)} Leading $\bar{p}$ spectra from the soft colour exchange models (SCI, GAL) and default {\sc Pythia} with varied remnant modeling. \label{fig:2}}
\vspace*{-3mm}
\end{figure}

CDF has also recently observed~\cite{cdf-dpe} dijets events with a leading antiproton and an opposite-side rapidity gap (called ``double pomeron exchange", DPE, based on the Regge interpretation). The $E_T$ and $\eta$ distributions of the jets as well as the ratio of these DPE to SD events 
are in good qualitative agreement with our SCI and GAL models. Having a gap or
leading particle on both sides in the DPE events means a stronger dependence on the details of the remnant treatment. Using the remnant splitting based on counting rules, the resulting cross section gets close to the measured one. 

In conclusion, the SCI and GAL models give a simple and reasonable description of rapidity gap events at both HERA and the Tevatron. They also reproduce the leading baryon data, although a remnant model dependence is found. Leading baryon data provide valuable additional information for the development of models for soft remnant interactions.

\vspace*{2mm}
We thank R.\ Enberg and K.\ Goulianos for very helpful discussions.

\end{document}